\documentclass{article}
\usepackage{spconf,amsmath,graphicx}

\usepackage{multirow}
\usepackage{hhline}
\usepackage{subfigure,graphicx}
\usepackage{pifont}
\usepackage{color}
\newcommand{\tabincell}[2]{\begin{tabular}{@{}#1@{}}#2\end{tabular}}

\newcommand{\xmark}{\ding{55}}
\newcommand{\cmark}{\ding{51}}

\title{Neural Architecture Search \\For LF-MMI Trained Time Delay Neural Networks}
\name{Shoukang Hu, Xurong Xie, Shansong Liu, Mingyu Cui, Mengzhe Geng, Xunying Liu, Helen Meng}
\address{The Chinese University of Hong Kong, Hong Kong SAR, China\\
\small \tt \{skhu, ssliu, mycui, mzgeng, xyliu, hmmeng\}@se.cuhk.edu.hk \small \tt \{xr.xie\}@link.cuhk.edu.hk}
\begin{document}
\ninept
\maketitle
\begin{abstract}
Deep neural networks (DNNs) based automatic speech recognition (ASR) systems are often designed using expert knowledge and empirical evaluation. In this paper, a range of neural architecture search (NAS) techniques are used to automatically learn two types of hyper-parameters of state-of-the-art factored time delay neural networks (TDNNs): i) the left and right splicing context offsets; and ii) the dimensionality of the bottleneck linear projection at each hidden layer. These include the DARTS method integrating architecture selection with lattice-free MMI (LF-MMI) TDNN training; Gumbel-Softmax and pipelined DARTS reducing the confusion over candidate architectures and improving the generalization of architecture selection; and Penalized DARTS incorporating resource constraints to adjust the trade-off between performance and system complexity. Parameter sharing among candidate architectures allows efficient search over up to $7^{28}$ different TDNN systems. Experiments conducted on the 300-hour Switchboard corpus suggest the auto-configured systems consistently outperform the baseline LF-MMI TDNN systems using manual network design or random architecture search after LHUC speaker adaptation and RNNLM rescoring. Absolute word error rate (WER) reductions up to 1.0\% and relative model size reduction of 28\% were obtained. Consistent performance improvements were also obtained on a UASpeech disordered speech recognition task using the proposed NAS approaches.
\end{abstract}
\begin{keywords}
Neural Architecture Search, Time Delay Neural Network, Speech Recognition
\end{keywords}

\section{Introduction}
Deep neural networks (DNNs) play a central role in state-of-the-art automatic speech recognition (ASR) systems\cite{waibel1989consonant, kingsbury2009lattice, vesely2013sequence, su2013error,  peddinti2015time, povey2016purely, povey2018semi}. When designing these systems, a set of DNN structure design decisions such as the hidden layer dimensionality and connectivity need to be made. These decisions are largely based on expert knowledge or empirical choice. As explicitly training and evaluating the performance of different network architectures is highly expensive, it is preferable to use automatic architecture design techniques~\cite{baluja1994reducing, kirschning1995parallel}. 

To this end, neural architecture search (NAS) approaches~\cite{ elsken2018neural} have gained increasing interests in recent years. The key objectives of NAS methods are three fold. First, it is crucial to produce an accurate performance ranking over different candidate neural architectures to allow the best system to be selected. Second, when operating at the same level of accuracy performance target, preference should be given to simpler architectures with fewer parameters in order to minimize the risk of overfitting to limited data. Furthermore, to ensure scalability and efficiency on large data sets, a search space containing all candidate systems of interest needs to be defined. 

Earlier forms of NAS techniques were based on neural evolution~\cite{stanley2002evolving}, where genetic algorithms were used to randomly select architecture choices at each iteration of mutation and crossover. Bayesian NAS methods based on Gaussian Process was proposed in~\cite{kandasamy2018neural}. Reinforcement learning (RL) based NAS approaches~\cite{zoph2016neural, pham2018efficient} have also been developed. In these techniques, explicit system training and evaluation are required. In addition, as the architecture hyper-parameters and actual DNN parameters are separately learned, e.g., within the RL controller and candidate systems, a tighter integration of both is preferred during NAS. 

Alternatively, differentiable architectural search (DARTS) techniques can be used~\cite{liu2018darts, xie2018snas, cai2018proxylessnas, hu2020dsnas, xie2020understanding}. Architectural search is performed over an over-parameterized parent super-network containing paths connecting all candidate DNN structures to be considered. The search is transformed into the estimation of the weights assigned to each candidate neural architecture within the super-network. The optimal architecture is obtained by pruning lower weighted paths. This allows both architecture selection and candidate DNN parameters to be consistently optimized within the same super-network model.

In contrast to the rapid development of NAS techniques in the machine learning and computer vision communities, there has been very limited research of applying these to speech recognition systems so far. In this paper, a range of DARTS based NAS techniques are used to automatically learn two architecture hyper-parameters that heavily affect the performance and model complexity of state-of-the-art factored time delay neural network (TDNN-F)~\cite{waibel1989consonant, peddinti2015time, povey2018semi, povey2016purely} acoustic models: i) the left and right splicing context offsets; and ii) the dimensionality of the bottleneck linear projection at each hidden layer. These include the standard DARTS method fully integrating the estimation of architecture weights and TDNN parameters in lattice-free Maximum Mutual Information (LF-MMI) training; Gumbel-Softmax DARTS that reduces the confusion between candidate architectures; pipelined DARTS that circumvents the overfitting of architecture weights using validation data; and penalized DARTS that further incorporates resource constraints to flexibly adjust the trade-off between performance and system complexity. Parameter sharing among candidate architectures was also used to facilitate efficient search over a large number of TDNN systems. Experiments conducted on a 300-hour Switchboard conversational telephone speech recognition task suggest the NAS configured TDNN-F systems consistently outperform the baseline LF-MMI trained TDNN-F systems using manually designed configurations. Absolute word error rate reductions up to 1.0\% and model size reduction of 28\% relative were obtained. In order to further evaluate the performance of the proposed NAS techniques, they were applied to automatically configure two sets of hyper-parameters of a state-of-the-art disordered speech recognition task based on the UASPEECH corpus~\cite{kim2008dysarthric}: skip connection between layers and dimensionality of factored TDNN weight matrices.

To the best of our knowledge, this paper is among the first to apply neural architecture search techniques to TDNNs in speech recognition tasks. In contrast, the vast majority of previous NAS research has been focused on computer vision applications~\cite{tan2019mixconv, Liu_2019_CVPR,chen2019detnas}. Existing NAS works in the speech community investigated non-TDNN based  architectures~\cite{veniat2019stochastic, mazzawi2019improving, li2019neural, chen2020darts, mo2020neural, kim2020evolved}. 

The rest of this paper is organized as follows. Section 2 presents a set of differentiable NAS techniques. Section 3 discusses the search space of TDNN-F models and necessary parameter sharing to improve search efficiency. Section 4 presents the experiments and results. Finally, the conclusions are drawn in Section 5.

\section{Neural Architecture Search}
\label{NAS_methods}

In this section, we present various forms of differentiable neural architecture search (DARTS) methods. With no loss of generality, we introduce the general form of DARTS architecture selection methods~\cite{liu2018darts, xie2018snas, cai2018proxylessnas, hu2020dsnas}. For example, the $l$-th layer output $\mathbf{h}^l$ can be computed as follows in the DARTS supernet:
\begin{equation}
\label{eq:darts_fw}
\setlength{\abovedisplayskip}{4pt}
\setlength{\belowdisplayskip}{4pt}
\begin{aligned}
\mathbf{h}^l=&\sum_{i=0}^{N^l-1}\!\lambda_i^l\phi_i^l(\mathbf{W}_i^{l}\mathbf{h}^{l-1})
\end{aligned}
\end{equation}
where $\lambda_i^l$ is the architecture weight for the $i$-th candidate choice in the $l$-th layer, $N^l$ is the total number of choices in this layer. The precise form of neural architectures being considered at this layer is determined by the linear transformation parameter $\mathbf{W}_i^{l}$ and activation function $\phi_i^l(\cdot)$ used by each candidate system. For example, when selecting the TDNN-F hidden layer context offsets, the linear transformation is a binary-valued matrix for each candidate architecture, and $\phi_i^l(\cdot)$ is represented by an identity matrix. When selecting the dimensionality of the bottleneck linear projection at each hidden layer, the linear transformation $\mathbf{W}_i^{l}=\widetilde{\mathbf{W}}_i^l\widehat{\mathbf{W}}_i^{l^T}$ is a decomposed matrix, while $\phi_i^l(\cdot)$ is also an identity matrix.

\vspace{1.0mm}\noindent
{\bf{2.1. Softmax DARTS:}}
The conventional DARTS~\cite{liu2018darts} system uses a Softmax function to model the architecture selection weight $\lambda_i^l$ as
\begin{equation}
\label{eq:softmax_fw}
\setlength{\abovedisplayskip}{4pt} 
\setlength{\belowdisplayskip}{4pt}
\begin{aligned}
\lambda_i^l=\frac{\exp(\log\alpha_i^l)}{\sum_{j=0}^{N^l-1}\exp(\log\alpha_j^l)}
\end{aligned}
\end{equation}

When using the standard back-propagation algorithm to update the architecture weights parameter $\lambda_i^l$, the loss function (including LF-MMI criterion~\cite{povey2016purely} considered in this paper) gradient against the $\lambda_i^l$ is computed as below.
\begin{equation}
\label{eq:softmax_bp}
\setlength{\abovedisplayskip}{4pt} 
\setlength{\belowdisplayskip}{4pt}
\begin{aligned}
\frac{\partial\mathcal L}{\partial\log\alpha_k^l}=\frac{\partial\mathcal{L}}{\partial\mathbf{h}^l}\sum_{i=0}^{N^l-1}\left(1_{i=k}\lambda_i^l-\lambda_i^{l}\lambda_k^{l}\right)\phi_i^l(\mathbf{W}_i^{l}\mathbf{h}^{l-1})
\end{aligned}
\end{equation}
where $1_{i=k}$ is the indicator function. When the DARTS supernet containing both architecture weights and normal DNN parameters is trained to convergence, including architecture parameters and normal DNN parameters, the optimal architecture can be obtained by pruning lower weighted architectures that are considered less important. However, when similar architecture weights are obtained using a flattened Softmax function, the confusion over different candidate systems increases and search errors may occur.

\vspace{1.0mm}\noindent
{\bf{2.2. Gumbel-Softmax DARTS:}}
In order to address the above issue, a Gumbel-Softmax distribution~\cite{maddison2016concrete} is used to sharpen the architecture weights to produce approximately a one-hot vector. This allows the confusion between different architectures to be minimised. The architecture weights are computed as, 
\begin{equation}
\label{eq:gumbel_reparam}
\setlength{\abovedisplayskip}{4pt} 
\setlength{\belowdisplayskip}{4pt}
\begin{aligned}
\lambda_{i}^{l}=\frac{\exp((\log\alpha_i^l+G_{i}^l)/T)}{\sum_{j=0}^{N^l-1}\exp((\log\alpha_j^l+G_{j}^l)/T)}
\end{aligned}
\end{equation}
where $G_{i}^l=-\log(-\log(U_{i}^l))$ is the Gumbel variable, and $U_{i}^l$ is a uniform random variable. When the temperature parameter $T$ approaches 0, it has been shown that the Gumbel-Softmax distribution is close to a categorical distribution~\cite{maddison2016concrete}.



Different samples of the uniform random variable $U_i^l$ lead to different values of $\lambda_i^l$ in Eq.~\ref{eq:gumbel_reparam}. The loss function gradient w.r.t $\log\alpha_k^l$ is computed as an average over $J$ samples of the architecture weights,
\begin{equation}
\label{eq:gumbel_bp}
\setlength{\abovedisplayskip}{4pt} 
\setlength{\belowdisplayskip}{4pt}
\begin{aligned}
\frac{\partial\mathcal{L}}{\partial\log\alpha_k^l}\!=\!\frac{1}{J}\!\sum_{j=0}^{J}\frac{\partial\mathcal L}{\partial\mathbf{h}^{l,j}}\!\sum_{i=0}^{N^l-1}\!\frac{1_{i=k}\lambda_{i}^{l,j}-\lambda_{i}^{l,j}\lambda_{k}^{l,j}}{T}\!\phi_i^l(\mathbf{W}_i^{l}\mathbf{h}^{l-1,j})
\end{aligned}
\end{equation}
where $\mathbf{\lambda}^{l,j}$ is the $j$-th sample weights vector drawn from the Gumbel-Softmax distrbution in the $l$-th layer, $\mathbf{h}^{l,j}$ is the output of $l$-th layer by using the $j$-th sample $\mathbf{\lambda}^{l,j}$. We assume the Gumbel-Softmax variables $\mathbf{\lambda}^{l}$ at different layers are mutually independent.

\vspace{1.0mm}\noindent
{\bf{2.3. Pipelined DARTS:}}
As both architecture weights and normal DNN parameters are learned at the same time in Softmax DARTS and Gumbel-Softmax DARTS systems, the search algorithms may prematurely select sub-optimal architectures at an early stage. Inspired by~\cite{guo2019single}, we decouple the update of normal DNN parameters and architecture weights into two separate stages performed in sequence. This leads to the pipelined DARTS approach. In order to prevent overfitting to the training data, a separate held-out data set taken out of the original training data is used. In Pipelined DARTS systems, the normal DNN parameters are updated to convergence on the training data first, while randomly sampled one-hot architecture weights drawn from a uniform distribution are used. In the following stage, we fix the normal DNN parameters estimated in the first stage in the supernet and update the architecture weights using the held-out data for both Softmax DARTS and Gumbel-Softmax DARTS. This 
produces the Pipelined Softmax DARTS (PipeSoftmax) and Pipelined Gumbel-softmax DARTS (PipeGumbel) systems.


\vspace{1.0mm}\noindent
{\bf{2.4. Penalized DARTS:}}
In order to flexibly adjust the trade-off between system performance and complexity, a penalized loss function incorporating the underlined neural network size is used. 
\begin{equation}
\label{eq:penalized_darts}
\setlength{\abovedisplayskip}{4pt} 
\setlength{\belowdisplayskip}{4pt}
\begin{aligned}
\mathcal{L} = \mathcal{L}_{LF-MMI} + \eta \sum_{l,i}\lambda_{i}^{l}C_{i}^l,
\end{aligned}
\end{equation}
where $C_{i}^l$ is the number of parameters of the $i$-th candidate considered at the $l$-th layer, and $\eta$ is the penalty scaling factor empirically set for different tasks.

\section{Search Space and Parameter Sharing}
This section describes the search space and its implementation when NAS methods of Sec.~\ref{NAS_methods} are used to automatically learn two sets of hyper-parameters of TDNN-F models: i) the left and right splicing context offsets; and ii) the dimensionality of the bottleneck linear projection at each hidden layer. Parameter sharing among candidate architectures used to facilitate efficient search over a large number of TDNN-F systems is also presented.

\vspace{1.0mm}\noindent
{\bf{3.1. TDNN-F Context Offset Search Space:}}
\label{layer_context_search_space}
\begin{figure}[t]
    \vspace{-4mm}
    \setlength{\abovecaptionskip}{-0cm}
    \setlength{\belowcaptionskip}{-0cm}
    \centering
    \includegraphics[width=2.9in]{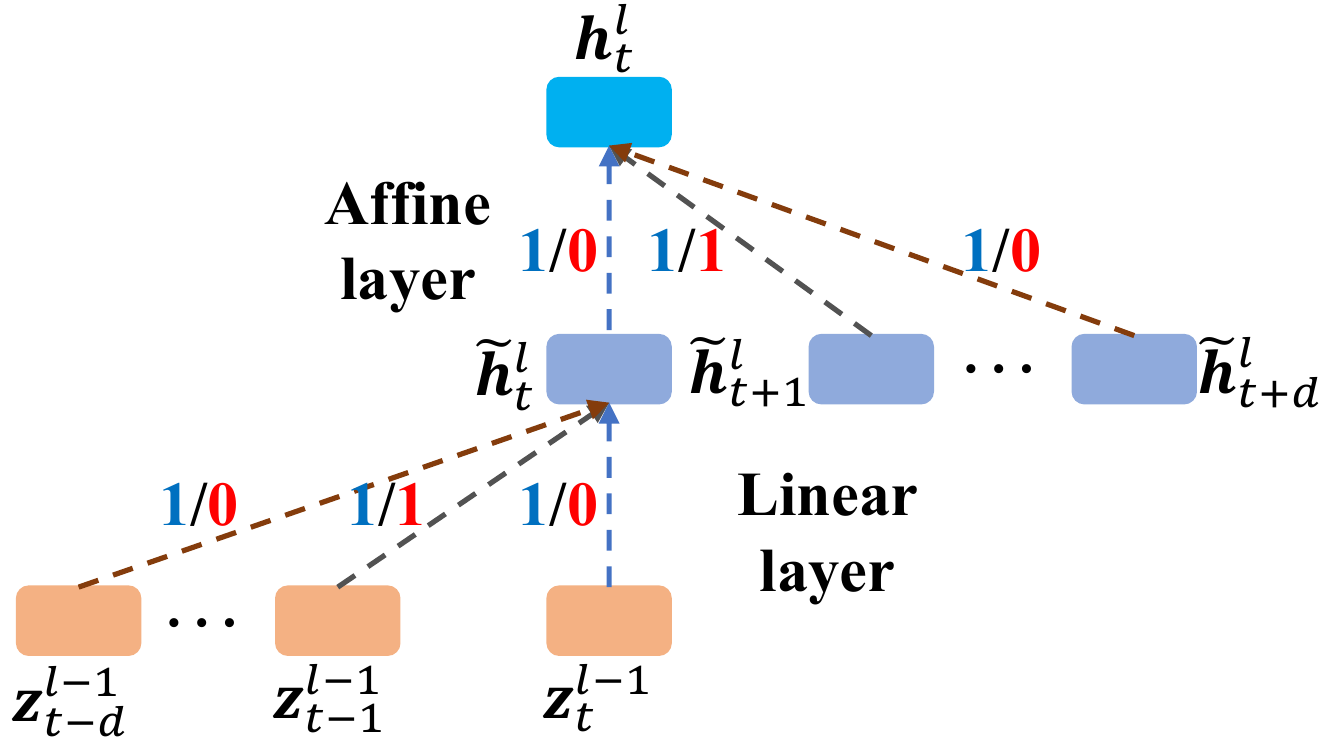}
    \caption{Example part of a supernet containing all the context offsets for a TDNN-F layer. Dashed lines with different colors represent different context choices in each linear (left context) and affine (right context) transforms. The blue integers denote the supernet system using all the context offsets, while the red integers represent a candidate offset choice of $\pm1$.}
    \label{fig:layer_context_search_space}
    \vspace{-5mm}
\end{figure}
Context offset settings play an important role in modeling the long temporal information in TDNN-F models. However, manually selecting context offsets is time-consuming for different applications. Inspired by the parameter-sharing used in earlier NAS research~\cite{pham2018efficient}, we design a TDNN-F supernet (Fig.~\ref{fig:layer_context_search_space}) to contain all possible choices of context offsets to the left (\{-d,0\}, $\cdots$, \{-1,0\}, \{0,0\}) and right (\{0,0\}, \{0,1\}, $\cdots$, \{0,d\}) at each layer during search. Note that {0,d} denote context offsets of 0 and d (right). For the supernet system, it requires the sparse context connection weights to be densely set as 1 for all context offsets. Any candidate TDNN-F model with particular context offsets, out of the total $(d+1)^{2L}$ possible choices, contained in the supernet is represented by setting the corresponding connection weights to be 1, while setting the others to be 0. 

\begin{figure}[h]
    \vspace{-4mm}
    \setlength{\abovecaptionskip}{-0cm}
    \centering
    \includegraphics[width=2.9in]{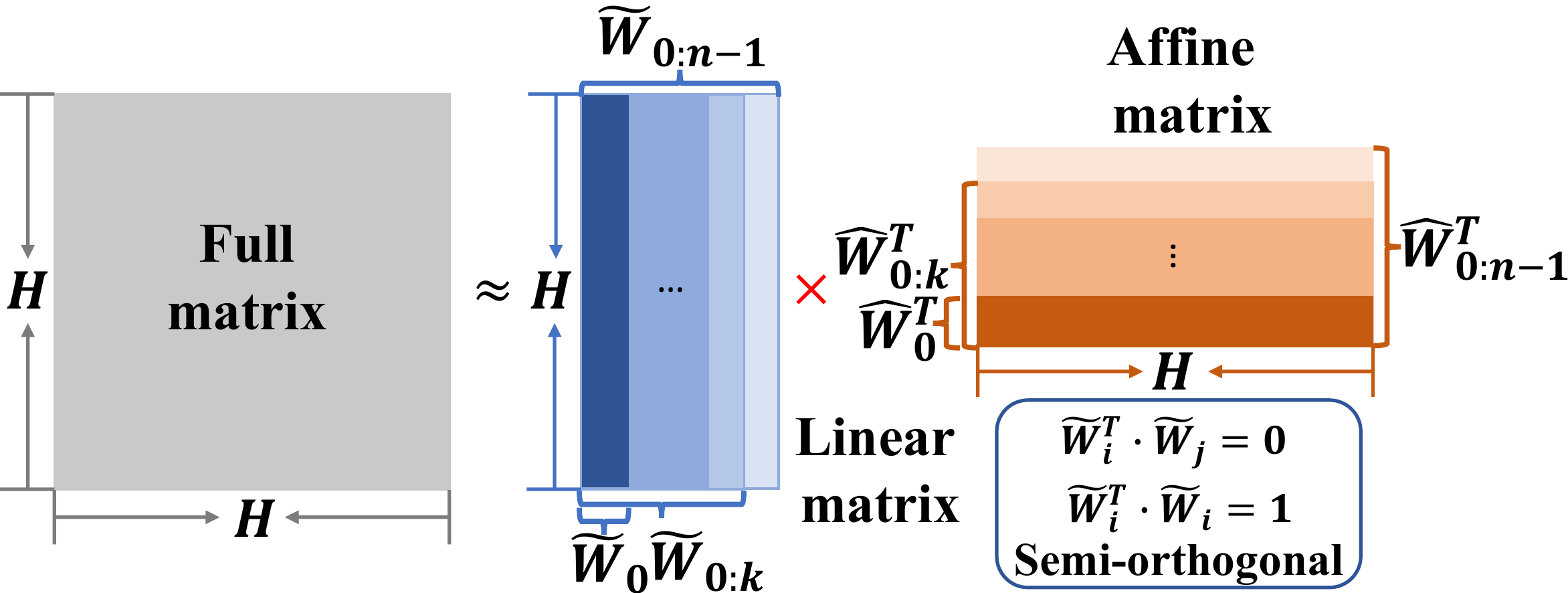}
    \caption{Example part of a supernet containing different bottleneck projection dimensionality choices in the TDNN-F hidden layer. The full weight matrix is factored into one semi-orthogonal linear weight matrix $\widetilde{\mathbf{W}}_{0:n-1}$ and one affine weight matrix $\widehat{\mathbf{W}}_{0:n-1}$. Architectures with different projection dimensions are represented by the corresponding submatrices starting form the first column.}
    \label{fig:layer_bottleneck_dim_search_space}
    \vspace{-4mm}
\end{figure}

\vspace{1.0mm}\noindent
{\bf{3.2. TDNN-F Bottleneck Dimensionality Search Space:}} Similarly, a TDNN-F supernet containing all the candidate TDNN-F with different projection dimensions is designed, as shown in Fig.~\ref{fig:layer_bottleneck_dim_search_space} for one hidden layer. When applying the NAS methods in Sec.2, $\phi_i^l(\cdot)$ of Eq. 1 is set as an identity matrix. In common with the standard TDNN-F model, the weight matrix $\mathbf{W}_i^{l}$ of $i$-th architecture choice in $l$-th layer is factored into one semi-orthogonal weight matrix $\widetilde{\mathbf{W}}^{l}_{0:n_i-1}$ and one affine weight matrix $\widehat{\mathbf{W}}^{l}_{0:n_i-1}$ as shown in Fig.~\ref{fig:layer_bottleneck_dim_search_space}. $n_i$ is the dimensionality of the $i$-th architecture. Parameter sharing among different candidate architectures' linear matrices $\widetilde{\mathbf{W}}_{0:k}$ (left and right from the first column) and affine matrices $\widehat{\mathbf{W}}_{0:k}$ (left and right from the first column) ($0\leq k\leq n-1$) is implemented by the corresponding submatrices extracted from the largest matrix $\widetilde{\mathbf{W}}_{0:n-1}$. Such sharing allows a large number of TDNN-F projection dimensionality choices at each of the 14 layers, e.g., selected from 8 values \{25, 50, 80, 100, 120, 160, 200, 240\} as considered in this paper, to be compared for selection during search. This leads to a total of $8^{14}$ candidate TDNN-F systems to be selected from.  

\section{Experimental Results}
This section presents our experiments carried out on the 300-hour Switchboard telephone speech recognition task using the Kaldi toolkit \cite{povey2011kaldi}. The GMM-HMM system~\cite{leggetter1995maximum, gales1998maximum}, TDNN-F acoustic model~\cite{povey2016purely}, and language model are similar to those described in~\cite{hu2019lf}. In the searching stage, TDNN-F supernet models are trained on the training set with one thread for 3 epochs, while architecture parameters of PipeSoftmax and PipeGumbel systems are updated for additional 3 epochs using a held-out data set by fixing the normal DNN parameters. Note that we randomly select 5\% of the original training set as the held-out data set and $T$ in the Gumbel-Softmax distribution is annealed from 1 to 0.03 in our experiments. Once candidate TDNN-F models are derived from the searching stage, they are trained for 3 epochs from scratch. Note that a matched pairs sentence-segment word error (MAPSSWE) based statistical significance test was performed at a significance level $\alpha=0.05$.
\vspace{1.0mm}\noindent
{\bf{4.1 TDNN-F Context Offset Search on SWBD:}}
\label{context_offset_search}
In this section, we describe the experimental results of searching context offsets at each layer by using various NAS methods of Sec.~\ref{NAS_methods}. In Table~\ref{context_offset_results}, systems (5)-(6), (9)-(10) perform the search over the total $4^{28}$ TDNN-F choices with maximum context offsets of $\pm3$ using 36 hours on a 32M param. supernet, while systems (7)-(8) perform the search over $7^{28}$ TDNN-F choices with the maximum context offsets of $\pm6$ using 60 hours on a 53M param. supernet. Two trends can be observed from Table~\ref{context_offset_results}. First, the Gumbel-Softmax DARTS system (Sys (8)) outperforms the baseline Kaldi recipe TDNN-F system (Sys (1)) by \textbf{0.3\%} and \textbf{1.0\%} absolute WER reductions on the \textbf{swbd} and \textbf{callhm} test sets. Even compared with additional manually designed TDNN-F systems (Sys (3)), the Gumbel-Softmax DARTS system (Sys 8) still produces \textbf{0.3\%} absolute WER reduction on the \textbf{callhm} test set. Second, Gumbel-Softmax DARTS systems (Sys (6), (8)) obtain better performance than Softmax DARTS systems (Sys (5), (7)). 

\begin{table}[h]
\vspace{-4mm}
\setlength{\abovecaptionskip}{-0cm}
\caption{Performance (WER\%) comparison of TDNN-F models (\#param:18M) configured with context offsets produced by the baseline system, manual designed systems, Softmax DARTS (Softmax), Gumbel-Softmax DARTS (Gumbel), Pipelined Softmax DARTS (PipeSoftmax), Pipelined Gumbel-Softmax DARTS (PipeGumbel) systems described in Sec.~\ref{NAS_methods}. $\{[a,b]\}$:$\{-c, d\}$ denotes context offsets $\{-c,0\}$ on the left and $\{0,d\}$ on the right used from $a$-th layer to $b$-th layer inclusive. $^\dagger$ denotes a statistically significant difference is obtained over the baseline system (Sys (1)).}
\label{context_offset_results}
\centering
\footnotesize
\setlength{\tabcolsep}{0.5mm}{
\begin{tabular}{c|c|c|cc}
\hline\hline
\multirow{2}*{Sys} & \multirow{2}*{Method} & \multirow{2}*{Context Offsets} & \multicolumn{2}{c}{Hub5' 00} \\
\cline{4-5}
~ & ~ & ~ & swbd & callhm \\
\hline\hline
1 & Baseline$^{\cite{povey2011kaldi}}$ & \{[1,3]\}:\{-1,1\}; \{4\}:\{0\}; \{[5,14]\}:\{-3,3\} & 10.0 & 20.8 \\
\hline\hline
2 & \multirow{3}*{Manual} & \{[1,3]\}:\{-1,1\}; \{4\}:\{0\}; \{[5,14]\}:\{-6,6\} & \textbf{9.8} & \textbf{20.1} \\
3 & ~ & \{[1,3]\}:\{-1,1\}; \{4\}:\{0\}; \{[5,14]\}:\{-9,9\} & \textbf{9.7} & \textbf{20.1} \\
4 & ~ & \{[1,3]\}:\{-1,1\}; \{4\}:\{0\}; \{[5,14]\}:\{-12,12\} & 10.1 & 20.8 \\
\hline\hline
5 & Softmax & \{1\}:\{0,1\};\{[2,14]\}:\{-3,3\} & 10.2 & 20.6 \\
\hline
6 & Gumbel & \{1,2\}:\{-2,2\}; \{3,4\}:\{-2,3\};\{[5,14]\}:\{-3,3\} & 9.9 & 20.1$^\dagger$ \\
\hline\hline
7 & Softmax & \tabincell{c}{\{1\}:\{0,1\};\{2\}:\{-6,1\};\\ \{3\}:\{-6,2\};\{[4,14]\}:\{-6,6\}} & 9.7 & 20.1$^\dagger$ \\
\hline
8 & Gumbel & \tabincell{c}{\{1,2\}:\{-3,2\}; \{3\}:\{-4,3\};\{4,5\}:\{-4,4\};\\ \{6,7\}:\{-4,6\};\{8\}:\{-5,6\};\{[9,14]\}:\{-6,6\}} & \textbf{9.7} & \textbf{19.8}$^\dagger$ \\
\hline\hline
9 & PipeSoftmax & \tabincell{c}{\{1\}:\{-1,3\};\{2,[4,12],14\}:\{-3,3\};\\ \{3\}:\{-2,3\};\{13\}:\{0,3\}} & 9.9 & 20.4$^\dagger$ \\
\hline
10 & PipeGumbel & \tabincell{c}{\{1\}:\{-1,3\};\{2,3\}:\{-2,3\};\{4\}:\{-3,2\};\\ \{5,6,[8,14]\}:\{-3,3\};\{7\}:\{0,3\}} & 9.9 & 20.4$^\dagger$ \\
\hline\hline
\end{tabular}}
\vspace{-4mm}
\end{table}

\vspace{1.0mm}\noindent
{\bf{4.2. TDNN-F Bottleneck Dimensionality Search on SWBD:}}
\label{bn_dim_search}
Table~\ref{bn_dim_results} shows the performance of searching bottleneck projection dimensions at each layer using NAS methods over the following 8 choices: \{25,50,80,100,120,160,200,240\}. This leads to a total of $8^{14}$ TDNN-F systems to be selected from. In comparison with the baseline Kaldi recipe~\cite{povey2011kaldi} TDNN-F system (Sys (1)), the Pipelined DARTS\footnote{Updating the architecture weights of Pipelined DARTS systems on the training data produces worse results than that using held-out data.} systems (Sys (11), (12)) with approximately the same number of parameters achieve comparable performance. If we further add the resource penalty to the objective loss function, the PipeGumbel system (Sys (14)) can produce \textbf{0.4\%} absolute WER reduction on the \textbf{callhm} test set and a relative model size reduction of \textbf{20\%} over the baseline Kaldi recipe~\cite{povey2011kaldi} TDNN-F system (Sys (1)), by selecting fewer bottleneck projection dimensions at higher layers. In addition, the Softmax DARTS system (Sys 9) does not outperform the baseline Kaldi recipe TDNN-F system (Sys (1)), which may be explained as prematurely selecting sub-optimal structures at an early stage when both architecture weights and normal DNN parameters are being trained before reaching convergence. Note that the supernet (24M parameters) training takes about 26 hours.
\begin{table}[t]
\vspace{-5mm}
\caption{Performance (WER\%, number of parameters) comparison of TDNN-F models configured with projection dimensions produced by the baseline system, manual designed systems, Softmax DARTS (Softmax), Gumbel-Softmax DARTS (Gumbel), Pipelined Softmax DARTS (PipeSoftmax), Pipelined Gumbel-Softmax DARTS (PipeGumbel) systems in Sec.~\ref{NAS_methods}. $\eta$ is the penalty factor in Eqn.~(\ref{eq:penalized_darts}). The dimensionality index denotes the index of 8 dimensionality choices: \{25,50,80,100,120,160,200,240\}. $^\dagger$ denotes a statistically significant difference is obtained over the baseline system (Sys (1)).}
\label{bn_dim_results}
\centering
\setlength{\abovecaptionskip}{-0cm}
\footnotesize
\setlength{\tabcolsep}{0.7mm}{
\begin{tabular}{c|c|c|c|cc|c|c}
\hline\hline
\multirow{2}*{Sys} & \multirow{2}*{Method} & \multirow{2}*{Bottleneck Dimension Id} & \multirow{2}*{$\eta$} & \multicolumn{2}{c|}{Hub5' 00}  & \#pa- & Ti-\\
\cline{5-6}
~ & ~ & ~ & ~ & swbd & callhm & ram & me\\
\hline\hline
1 & Baseline$^{\cite{povey2011kaldi}}$ & 5 5 5 5 5 5 5 5 5 5 5 5 5 5 & - & 10.0 & 20.8 & 18M & 18h\\
\hline\hline
2 & \multirow{7}*{Manual} & 0 0 0 0 0 0 0 0 0 0 0 0 0 0 &\multirow{7}*{-} & 11.2 & 23.5 & 7M & 10h \\
3 & ~ & 1 1 1 1 1 1 1 1 1 1 1 1 1 1 & & 10.3 & 21.6 &  9M & 14h\\
4 & ~ & 2 2 2 2 2 2 2 2 2 2 2 2 2 2 & & 10.2 & 21.1 &  11M & 15h\\
5 & ~ & 3 3 3 3 3 3 3 3 3 3 3 3 3 3 & & 10.0 & 20.6 & 13M & 16h\\
6 & ~ & 4 4 4 4 4 4 4 4 4 4 4 4 4 4 & & 10.0 & 20.7 & 15M & 17h\\
7 & ~ & 6 6 6 6 6 6 6 6 6 6 6 6 6 6 & & 10.2 & 20.1 & 21M & 19h\\
8 & ~ & 7 7 7 7 7 7 7 7 7 7 7 7 7 7 & & 10.1 & 20.5 & 24M & 20h\\
\hline\hline
9 & Softmax & 0 0 0 0 0 0 7 7 7 7 7 7 7 7 & \multirow{2}*{0} & 10.2 & 21.3 & 17M & 18h\\
\cline{1-3}\cline{5-7}
10 & Gumbel & 3 3 4 0 5 5 6 6 5 6 5 5 5 5 & & 10.1 & 20.5 & 17M & 18h\\
\hline\hline
11 & PipeSoftmax & 7 6 6 4 7 6 4 6 6 0 0 0 0 7 & \multirow{2}*{0} & 10.2 & 20.5 & 17M & 18h\\
\cline{1-3}\cline{5-7}
12 & PipeGumbel & 6 6 6 3 7 6 4 4 4 6 0 0 7 7 & & 9.9 & 20.7 & 18M & 18h\\
\hline\hline
13 & PipeSoftmax& 5 3 4 0 6 5 5 5 5 0 0 0 0 7 & .02 & 10.2 & 21.0 & 14.1M & 17h\\
\hline
14 & PipeGumbel & 5 3 4 0 4 5 4 3 4 3 1 2 6 6 & .03 & \textbf{9.9} & \textbf{20.4}$^\dagger$ & \textbf{14.6M} & 17h\\
\hline\hline
\end{tabular}}
\vspace{-5mm}
\end{table}

\vspace{1.0mm}\noindent
{\bf{4.3. Context Offsets \& Projection Dims Search on SWBD:}}
Performance comparison of searching both context offsets and bottleneck projection dimensionality using NAS and Random Search (selecting the best performed model from 5 randomly sampled models) is shown in Table~\ref{context_offset_bn_dim_results}. Two main trends can be observed. First, by searching the context offsets using Gumbel-Softmax DARTS (first stage of PipeGumbel) and projection dimensions using PipeGumbel DARTS in the same supernet, the system (3)\footnote{With projection dimensions: \{160,100,100,100,120,100,80,120,50,80,80 ,80,100,120\} and context configurations: \{1\}:\{-2,2\};\{2\}:\{-2,4\};\{3,4\}:\{-3,3\};\{5\}:\{-3,2\};\{6\}:\{-3,4\};\{7\}:\{-4,4\};\{8\}:\{-4,6\};\{[9,14]\}:\{-6,6\}. The supernet (76M parameters) training takes about 72 hours on a NVIDIA P100 GPU card, while the derived candidate model training takes about 20 hours.} in Table~\ref{context_offset_bn_dim_results} produces \textbf{0.4\%}- \textbf{1.0\%} absolute WER reductions on three test sets and a relative model size reduction of \textbf{28\%} over the baseline Kaldi recipe TDNN-F system (Sys (1)). The system (3) also outperforms the Random Search system (Sys (2)). Second, significant performance improvements can still be retained after learning hidden unit contribution (LHUC) based speaker adaptation and recurrent neural network language model (RNNLM) rescoring were performed.



\begin{table}[h]
\vspace{-5mm}
\caption{Performance (WER\%, number of parameters) of TDNN-F baseline systems, TDNN-F models configured with both context offsets and projection dimensions produced by NAS or Random Search before and after applying LHUC speaker adaptation and RNNLM rescoring. NAS denotes that offsets and dimensions are searched using the Gumbel-Softmax techniques in the same supernet, while Random Search selects the best performed model from 5 randomly sampled models. $^\dagger$ denotes a statistically significant difference is obtained over the baseline system (Sys (1), Sys (4)).}
\label{context_offset_bn_dim_results}
\setlength{\abovecaptionskip}{-0cm}
\centering
\footnotesize
\setlength{\tabcolsep}{0.5mm}{
\begin{tabular}{c|c|c|c|c|cc|c|c|c}
\hline\hline
\multirow{2}*{Sys} & \multirow{2}*{Method}&
\multirow{2}*{LHUC} & \multirow{2}*{LM} & \multirow{2}*{$\eta$} & \multicolumn{2}{c|}{Hub5' 00} & \multirow{2}*{Rt03S} & \multirow{2}*{Rt02} & \multirow{2}*{\#param} \\
\cline{6-7}
~ & ~ & ~ & & & swbd & callhm & ~ & ~ & ~\\
\hline\hline
1 & Baseline$^{\cite{povey2011kaldi}}$ & \multirow{3}*{\xmark} & \multirow{3}*{4g} & - & 10.0 & 20.8 & 18.1 & 17.4 & 18M\\
\cline{1-2}\cline{5-10}
2 & Manual & ~ & ~ & - & 9.7 & 20.3$^\dagger$ & \textbf{17.2}$^\dagger$ & \textbf{16.7}$^\dagger$ & 13M\\
\cline{1-2}\cline{5-10}
3 & Random Search & ~ & ~ & - & 9.8 & 21.2 & 17.6 & 17.3 & 14M\\
\cline{1-2}\cline{5-10}
4 & NAS & ~ & ~ & 0.03  & \textbf{9.6}$^\dagger$ & \textbf{19.8}$^\dagger$ & \textbf{17.2}$^\dagger$ & \textbf{16.7}$^\dagger$ & \textbf{13M}\\
\hline\hline
5 & Baseline$^{\cite{povey2011kaldi}}$ & \multirow{3}*{\cmark} & \multirow{3}*{+RNN} & - & 8.2 & 17.5 & 14.8 & 14.3 & 18M\\
\cline{1-2}\cline{5-10}
6 & Manual & ~ & ~ & - & 8.1 & 17.2$^\dagger$ & 14.4$^\dagger$ & 13.9$^\dagger$ & 13M\\
\cline{1-2}\cline{5-10}
7 & Random Search & ~ & ~ & - & 8.2 & 17.9 & 14.9 & 14.4 & 14M\\
\cline{1-2}\cline{5-10}
8 & NAS & ~ & ~ & 0.03  & 8.1 & \textbf{16.9}$^\dagger$ & \textbf{14.3}$^\dagger$ & \textbf{13.8}$^\dagger$ & \textbf{13M}\\
\hline\hline
\end{tabular}}
\vspace{-5mm}
\end{table}
\vspace{1.0mm}\noindent
{\bf{4.4. NAS Experiments on UASPEECH:}} To further evaluate the performance of the proposed NAS techniques, they were applied to configure two sets of hyper-parameters of a state-of-the-art disordered speech task based on the UASPEECH corpus~\cite{kim2008dysarthric}: skip connection between hidden layers and projection dimensionality of factored DNN weight matrices (similar to TDNN-F in Sec. 4.2). The detailed description of the baseline system can be found in~\cite{liu2020exploiting, geng2020investigation}. The results in Table~\ref{ua_results} suggest that NAS techniques\footnote{With projection dimensionality \{160,160,160,120,120\} and no skip connection at each layer.} applying the PipeGumbel DARTS consistently outperform the baseline systems.

\vspace{-5mm}
\begin{table}[h]
\caption{Performance (WER\%, number of parameters) of baseline DNN systems, models configured with the skip connections and projection dimensions produced by NAS or Random Search methods. NAS searches skip connections and dimensions using the PipeGumbel techniques in the same supernet, while Random Search selects the best performed model from 5 randomly sampled models. $^\dagger$ denotes a statistically significant difference is obtained over the baseline system (Sys (1)).}
\label{ua_results}
\centering
\begin{tabular}{c|c|c|c|c}
\hline\hline
Sys & Method  & $\eta$ & Test & \#param \\
\hline\hline
1 & Baseline (Manual)  & - & 31.45 & 5.9M \\
\hline
2 & Random Search  &  - & 32.30 & 5.16M \\
\hline
3 & NAS & 0.21 & \textbf{30.83}$^\dagger$ & \textbf{4.7M}\\
\hline\hline
\end{tabular}
\end{table}
\vspace{-5mm}
\section{Conclusions}
\vspace{-1.5mm}
In this paper, a range of neural architecture search (NAS) techniques is investigated to automatically learn three hyper-parameters that heavily affect the performance and model complexity of state-of-the-art factored time delay neural network (TDNN-F) acoustic models: i) the left and right splicing context offsets; ii) the dimensionality of the bottleneck linear projection. Experimental results obtained from Switchboard and UASPEECH suggest NAS techniques can be used for the automatic configuration of DNN based speech recognition systems and allow their wider application to different tasks.
\vspace{-2mm}
\section{Acknowledgements}
\vspace{-2mm}
This research is supported by Hong Kong RGC GRF No. 14200220, Theme-based Research Scheme T45-407/19N, ITF grant No. ITS/254/19, and SHIAE grant No. MMT-p1-19.
\vspace{-2mm}
\bibliographystyle{IEEEbib}
\bibliography{mybib}




\end{document}